**Real-time detection of critical slowing-down at the superconducting phase transition**

Guillermo Nava Antonio,[1] Théo Courtois,[2] Corentin Pfaff,[2] KM Shivangi Shukla,[1] Asle Sudbø,[3] Stéphane Mangin,[2] Thomas Hauet,[2] and Chiara Ciccarelli[1,3,*]

[1]Cavendish Laboratory, University of Cambridge, Cambridge, United Kingdom
[2]Institut Jean Lamour, Université de Lorraine, Nancy, France
[3]Center for Quantum Spintronics, Department of Physics, Norwegian University of Science and Technology, Trondheim, Norway

*Contact author: cc538@cam.ac.uk

**ABSTRACT**. We employ optical pump-THz probe spectroscopy to chart the ultrafast superconductivity suppression in NbN over a broad range of excitation fluences. Our measurements uncover a pronounced lengthening of the superconductivity quenching time when the absorbed optical energy is close to the condensation energy of the superconductor, which constitutes a non-equilibrium analog of critical slowing-down on a timescale comparable to that of superconducting fluctuations. Time-dependent Ginzburg-Landau simulations reproduce this behavior and ascribe it to the flattening of the free energy landscape at the dynamical phase transition boundary. Our findings represent a direct observation in real time of slowed-down superconductivity dynamics in proximity to a critical point and open a pathway for investigating out-of-equilibrium critical phenomena with time-resolved THz spectroscopy.

The excitation of a superconductor with an ultrashort (femtosecond) optical pulse results in a transient energy-mode imbalance in the quasiparticle (QP) spectrum and a concomitant non-thermal reduction of the superconducting gap ($\Delta$) [1–3]. The study of such photo-induced dynamics has shed light into the low-energy electronic structure [4–6] and mechanisms behind Cooper pair formation and breaking [7–9] in conventional and high-temperature superconductors, while also contributing to the development of ultrafast radiation detectors [10].

In more detail, the absorption of ultrashort optical pulses produces a QP avalanche near the gap edge and a non-equilibrium population of high-frequency phonons (HFPs) with energies greater than $2\Delta$ [11]. These two sub-systems exchange energy through Cooper pair breaking and recombination processes, whose cyclical occurrence gives rise to a suppression of the superconducting order lasting a few picoseconds in BCS systems [12]. Typically, the rebuilding of superconductivity is hampered by the slow dissipation of HFPs by anharmonic decay or escape to the surrounding environment, leading to a sub-nanosecond recovery in conventional superconductors [5]. In the weak perturbation regime, these out-of-equilibrium dynamics have been widely studied in pump-probe experiments [3,13] and elucidated in terms of the phonon bottleneck effect through the Rothwarf-Taylor (RT) model [14,15].

Nevertheless, the dynamical response of superconductors excited by strong optical pulses capable of destroying the Cooper pair condensate remains comparatively unexplored. Pioneering studies have established the characteristic optical energy required to non-adiabatically quench the superconducting phase and its difference from the energy needed to heat up the superconductor above the critical temperature ($T_c$) [1,16,17]. In this work, we extend this line of research by investigating the timescale of photo-induced superconductivity dynamics when the gap is transiently closed.

Our experiments reveal an elongation of the dynamics closely related to the phenomenon of critical slowing-down. The latter describes the substantial increase of the relaxation time following a small perturbation as a critical point is approached [18]. Beyond condensed matter, this effect is ubiquitous in systems with critical transitions [19], arising in fields as varied as medicine [20] or climate science [21]. In superconductors, critical slowing-down can be manifested as an enhancement of the microwave conductivity [22,23], the freezing of the vortex fluid in the mixed state [24,25], or a strengthening of magnetic fluctuations [26,27]. Real-time measurements of the superconductivity recovery after optical excitation in $Bi_2Sr_2CaCu_2O_{8+\delta}$ found scaling behavior and a considerable rise of the scaling time constant near $T_c$ [28]. However, a conclusive identification of critical slowing-down in those measurements is hindered due to the simultaneous presence of the phonon bottleneck effect, which also significantly decelerates the recovery close to $T_c$ [29].

In this Letter, we present real-time experiments demonstrating a far-from-equilibrium counterpart of critical-slowing down in the laser-induced superconductivity quenching of the BCS system NbN. By means of optical pump-THz probe (OPTP) spectroscopy, we identify a distinct prolongation of the superconductivity suppression when the system is transiently taken to the phase transition boundary. The application of the time-dependent Ginzburg-Landau theory qualitatively reproduces our results and highlights their connection with critically slowed down kinetics in the vicinity of equilibrium.

Our work is centered on a 15-nm-thick NbN film grown via magnetron sputtering as described in Appendix A. The emergence of superconductivity in the film was confirmed through vibrating-sample magnetometry (see Fig. S1 in [30]). The equilibrium electrodynamic properties of NbN were examined through THz time-domain spectroscopy (THz TDS) [31] and its superconductivity suppression dynamics were studied through OPTP experiments [32,33]. In the latter, 35-fs-long laser pulses induced a non-equilibrium state in the superconductor that was monitored by transmitting THz pulses at an adjustable time delay $t$ after the laser pump arrival. More precisely, we measured the change in the THz signal due to the optical perturbation: $\Delta S(t_{\text{THz}}, t) = S_{\text{OFF}}(t_{\text{THz}}) - S_{\text{ON}}(t_{\text{THz}}, t)$, where $S_{\text{ON}}$ and $S_{\text{OFF}}$ are the THz transmission with and without the influence of the pump, respectively, and $t_{\text{THz}}$ corresponds to the time axis associated with the electro-optic detection of THz radiation (see Appendix B for further details).

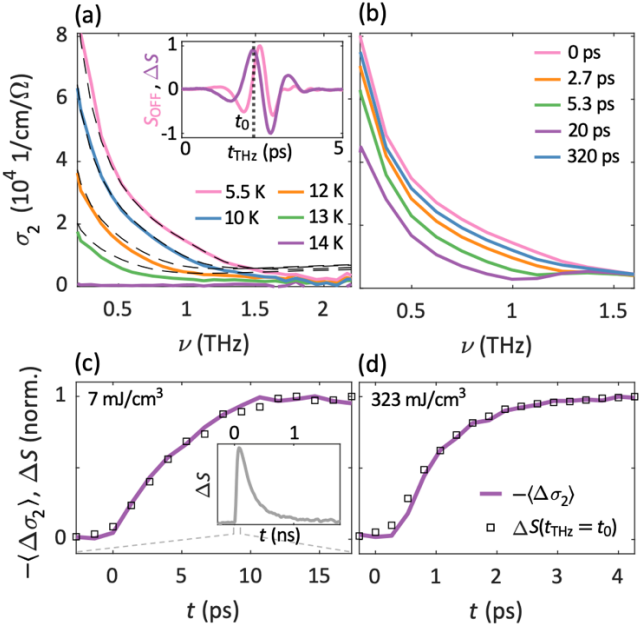

FIG. 1. Conductivity of the NbN film. (a) Imaginary part of the conductivity at equilibrium. The solid and dashed lines correspond to experiments and theory, respectively. Inset: THz transmission at 5.5 K ($S_{OFF}$) and its optically induced change ($\Delta S$) for $t = 20$ ps in arbitrary units. (b) Imaginary conductivity at different moments after the pump arrival measured at 5.5 K with $\Omega = 23$ mJ/cm³. (c) and (d) Dynamics of the photo-induced conductivity compared with the OPTP signal obtained at 5.5 K with $\Omega = 7$ and 323 mJ/cm³. Each time series is normalized to unity. Inset: OPTP signal exhibiting the superconductivity suppression and recovery.

First, we characterized the temperature-driven superconducting phase transition using THz TDS. A representative example of a transmitted THz pulse ($S_{OFF}$) is shown in the inset of Fig. 1(a). Such THz waveforms exhibited an abrupt change below $T_c$ stemming from the opening of the superconducting gap [34] (see Fig. S2 [30]). By Fourier transformation of the experimental data and application of Tinkham's formula for electromagnetic transmission [35,36], we obtained the NbN conductivity $\sigma_{NbN} = \sigma_1 + i\sigma_2$. In our analysis, we focus on the imaginary part, shown in Fig. 1(a) as a function of frequency ($\nu$), because of its relevance to the OPTP measurements, which is discussed below. The real part is reported in Fig. S2 [30]. In conformity with the BCS theory [34], $\sigma_2$ displayed a peak at $\nu = 0$ Hz associated with the inductive response of the Cooper pair condensate. Moreover, $\sigma_1$ was largely diminished for $\nu \lesssim 2\Delta/h$, where $h$ is Planck's constant, owing to the gap in the QP density of states.

In order to extract the superconducting gap, we modeled the experimental results with the theory of Zimmermann et al. [37] as delineated in Supplemental Note 1 [30]. The calculation, represented by the dashed lines in Fig. 1(a), reasonably reproduced the experiment by considering a QP scattering time of 55 fs, an average (across the sample) zero-temperature gap $\Delta_0$ of 0.62 THz, and an average critical temperature of 13.5 K—the latter value being in good agreement with the magnetometry data in Fig. S1.

Next, we utilized OPTP spectroscopy to examine the out-of-equilibrium state produced by an ultrashort optical perturbation. Specifically, from measurements of $\Delta S$, we determined $\sigma_{NbN}$ at different instants following the excitation [5,33]. As depicted in Fig. 1(b), the pair-breaking effect of the pump resulted in a progressive weakening of the low-frequency peak of $\sigma_2$ in the first few picoseconds after the laser pulse arrival, corresponding to the suppression of superconductivity. On a longer timescale of the order of hundreds of picoseconds, $\sigma_2$ was partially restored to its equilibrium value, indicating the recovery of superconductivity. Similar behavior has been observed in other conventional superconductors, such as $Nb_3Sn$ [12] and $MgB_2$ [5,16], as well as in cuprates [38–40].

Qualitatively, the effect of the laser pump is akin to increasing the ambient temperature [9,12]. However, discrepancies exist between the detailed frequency dependence of $\sigma_{NbN}$ in the two cases since the photo-induced changes of the conductivity are connected with a non-equilibrium QP distribution that, in general, deviates from the Fermi-Dirac function [41]. This is most clearly seen in the hump around $2\Delta/h \approx 1.2$ THz in $\sigma_1$ for $t = 20$ ps, which is not present at equilibrium (see Figs. S2(c) and S3(a) [30]).

In an alternative experimental approach, we measured $\Delta S$ at a fixed point $t_{THz} = t_0$ and as a function of pump delay $t$, a quantity we will henceforth refer to as the OPTP signal. This methodology entailed a shorter data acquisition time and, therefore, was more resistant to laser drift than the measurement of full THz pulses. As illustrated in the inset of Fig. 1(a), $t_{THz} = t_0$ corresponded to the maximum of $|\Delta S|$, i.e., the point of the transmitted waveforms that changed the most upon optical pumping.

In the weak perturbation regime, the OPTP signal has been shown to be proportional to the temporal evolution of the photo-induced conductivity modulation at any given frequency [5,16,17] (see also Fig S3): $\Delta S(t_0, t) \propto \Delta\sigma_j(\nu, t) = \sigma_j(\nu, t) - \sigma_j(\nu, t \to \infty)$, where the last term is the equilibrium conductivity and $j = 1$ or 2. Thus, the average of $\Delta\sigma_2(\nu, t)$ over $\nu$ also mimics the pump delay dependence of $\Delta S(t_0, t)$, as evinced by Fig 1(c). There, $\langle\Delta\sigma_2(t)\rangle = \int_{-\infty}^{\infty} \Delta\sigma_2(\nu, t) W(\nu) d\nu$ where $W(\nu)$ corresponds to the bandwidth of the OPTP experiments and is reported in the inset of Fig. S3(d); we employed such a weighted mean due to the lowering of the signal-to-noise ratio with distance from the center of the bandwidth.

As shown in Fig. 1(d), we verified that the relationship between $\langle\Delta\sigma_2\rangle$ and $\Delta S$ was preserved in the case of intense

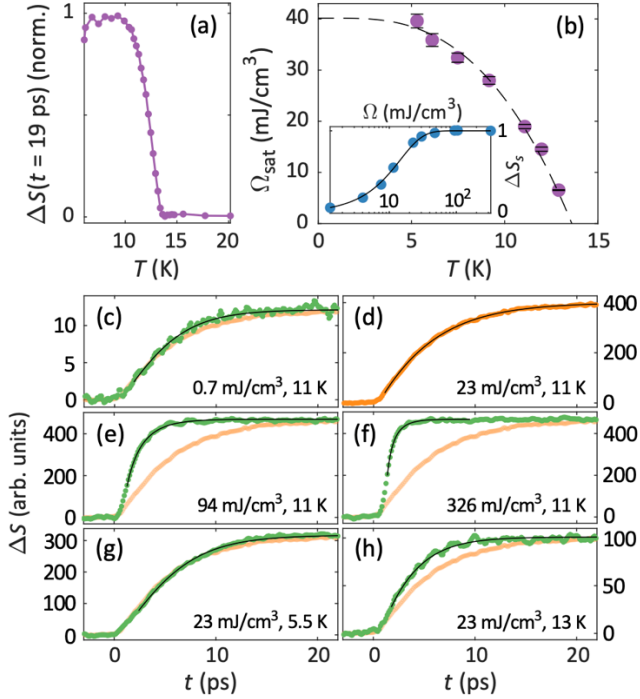

FIG. 2. OPTP response during the superconductivity suppression. (a) OPTP signal for $\Omega = 23$ mJ/cm$^3$ and $t = 19$ ps—a pump delay roughly at the end of the superconductivity quenching. (b) Saturation energy density determined from the fluence dependence of $\Delta S_s$, as illustrated in the inset for $T = 11$ K. The dashed line is a fit to a constant times $\Delta^2(T)$, with the latter calculated as in Fig. S2(b). (c)-(h) OPTP transients measured at selected $T$ and $\Omega$. For comparison, the orange data points, which correspond to a location in the $T, \Omega$ phase space close to the slowed down region, are rescaled and repeated in all panels. The solid lines are fits to Eq. (2).

excitations (larger than $\Omega_{sat}$ defined below). Since the imaginary part of the conductivity is a measure of the Cooper pair density [2,38], the OPTP signal thus provided a simple way of monitoring the changes in the strength of the superconducting condensate. In particular, the rise of $\Delta S$ in Figs. 1(c) and (d) indicated the quenching of superconductivity, while the recovery is denoted by the subsequent decay of $\Delta S$ (see inset of Fig. 1(c) and Supplemental Note 2 [30]). Moreover, as seen in Fig. 2(a), the OPTP signal vanished at $T_c$, confirming that $\Delta S$ was only sensitive to the superconducting phase in contrast to optical probe experiments that yield a non-zero response in the normal metallic state [7,11,42]

We took advantage of this measurement scheme to survey the superconductivity suppression dynamics in a wide range of temperatures and optical fluences. Figures 2(c)-(h) reveal that both the timescale and magnitude of the superconductivity quenching varied with the aforementioned experimental conditions. In the inset of Fig. 2(b), we plot the amplitude of the OPTP signal at the end of the superconductivity suppression process ($\Delta S_s$) at a fixed temperature and as a function of absorbed laser fluence ($\Omega$).

The latter quantity was calculated using the transfer matrix method of optics [43]. For weak excitations, $\Delta S_s$ linearly increased with $\Omega$. At sufficiently high energy densities, $\Delta S_s$ saturated, suggesting a complete collapse of the Cooper pair condensate. We extracted the saturation fluence ($\Omega_{sat}$) by fitting these data with the empirical model $\Delta S_s \propto \tanh(\Omega/\Omega_{sat})$ (solid line in the inset of Fig. 2(b)). The temperature dependence of $\Omega_{sat}$ is given in the main panel of Fig. 2(b). Previous works have found that, in conventional superconductors, the quantity $\Omega_{sat}$ is related to the superconductivity condensation energy [9,16,17]. Indeed, our $\Omega_{sat}$ data (e.g., $40 \pm 1$ mJ/cm$^3$ at 5.5 K, the base temperature of our cryostat) are commensurate with the condensation energy of NbN: $E_c = N(0)\Delta_0^2/2 = 22$ mJ/cm$^3$ at zero temperature, where $N(0) = 0.44$ eV$^{-1}$ spin$^{-1}$ formula unit$^{-1}$ is the density of states at the Fermi level [9]. Furthermore, Fig. 2(b) evidences that $\Omega_{sat}(T)$ scales with $\Delta^2(T)$. These observations support the interpretation that $\Omega_{sat}$ is the characteristic laser energy required to fully destroy the superconducting phase. As we will shortly show, this is also the energy scale that governs the speed of the photo-induced nonequilibrium dynamics.

We gauged the timescale of the superconductivity suppression process by fitting the OPTP transients as explained in Appendix C (solid lines in Figs. 2(c)-(h)), which yielded the characteristic suppression time $\tau_s$. Figure 3(a) displays the obtained $\tau_s$ values at 5.5 K and 12 K as a function of absorbed laser energy density. At low temperatures ($T \lesssim 0.5T_c$) and in the weak perturbation limit ($\Omega \ll \Omega_{sat}(T)$), the suppression time grew with decreasing fluence, in quantitative agreement with the RT model (see black curve in Fig. 3(a) and Appendix C). This lengthening of $\tau_s$ originated from an inhibition of Cooper pair breaking through HFP absorption caused by the relatively small HFP population excited by a low laser intensity [15]. For intermediate fluences beyond the validity of the RT theory, we observed a peak in $\tau_s$ close to $\Omega_{sat}(T)$ (dashed lines in Fig. 3(a)) at all studied temperatures, which we expand upon below. For stronger perturbations, the suppression time monotonically declined as $\Omega$ was increased.

Figure 3(b) shows how the suppression time varied with temperature for different fixed absorbed fluences. Analogously to Fig. 3(a), the dashed lines in panel (b) mark the location of the temperature $T_{sat}$ that satisfied $\Omega_0 = \Omega_{sat}(T_{sat})$ for the constant fluence $\Omega_0$ with which each data group was measured. In the low-$\Omega_0$ regime, $\tau_s$ sharply rose as $T_c$ was approached, which is consistent with the vanishing of the Cooper pair breaking rate at $T_c$ predicted by near-equilibrium calculations [41,44]. More generally, $\tau_s$ peaked when the experimental conditions were such that the optical pump almost eradicated the superconducting condensate.

An extended data set summarizing our $\tau_s$ measurements is presented in Fig. 3(c) (see also the equivalent one-dimensional plots in Fig. S5 [30]). In accordance with the remarks above, the color map indicates that $\tau_s$ had local

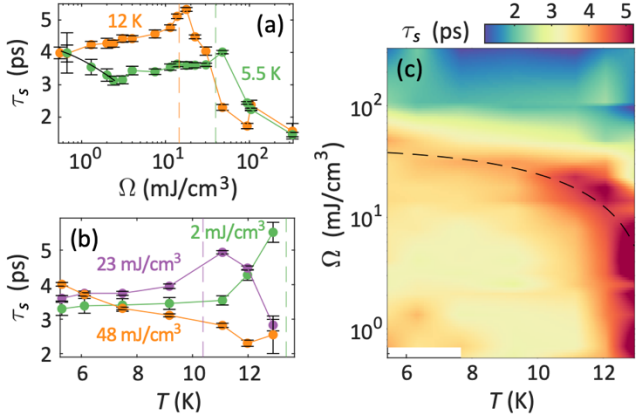

FIG. 3. Superconductivity suppression time. (a) Fluence dependence of $\tau_s$ at 5.5 and 12 K. The dashed lines indicate $\Omega_{\text{sat}}(T)$ at those temperatures. The solid curve is a fit to the RT model. (b) Temperature dependence of $\tau_s$ measured with the fixed fluences $\Omega_0 = 2$, 23, and 48 mJ/cm$^3$. The dashed lines denote $T_{\text{sat}}$. (c) Suppression time as a function of temperature and fluence. The dashed line is the same as in Fig. 2(b), i.e., it represents $\Omega_{\text{sat}}(T)$.

maxima—as a function of both temperature and fluence—around the locus $\Omega = \Omega_{\text{sat}}(T)$, which is depicted by the dashed line in Fig. 3(c). Such a prolongation of the superconductivity quenching process is clearly visible in the raw OPTP signal, as highlighted by the orange data points in Figs. 2(c)-(h). As argued next, we interpret this protraction as an analog of the phenomenon of critical slowing-down for a strongly out-of-equilibrium superconducting system.

With the aim of elucidating this parallelism, we modeled our results with the time-dependent Ginzburg-Landau theory of continuous phase transitions. The latter describes the temporal evolution of the order parameter ($\psi$) through the expression $\partial\psi/\partial t = -\Gamma\, \delta L/\delta\psi$, where $\Gamma$ is a proportionality constant and $L$ the free energy [18]. Typically, this formalism is employed in the vicinity of equilibrium [45,46]. Nonetheless, inspired by its successful application in the picosecond photo-induced charge density wave transition in various systems [47,48], we used the Ginzburg-Landau phenomenology to analyze the far-from-equilibrium superconductivity suppression dynamics in NbN. Crucially, the Ginzburg-Landau framework permits the variations of the free energy and of the order parameter to occur on different timescales, as in our non-adiabatic experiments.

In our calculations, we considered the free energy of a homogenous system in the absence of an electromagnetic field and under a rigid gauge to ensure that $\psi \in \mathbb{R}$ [49]:

$$L(t,T) = a(t,T)\psi^2 + (b/2)\psi^4, \quad (1)$$

$$a(t,T) = \hat{a}(T-T_c) + \theta(t)k\Omega e^{-t/\tau_0},$$

where $b$, $\hat{a}$ and $k$ are positive constants and $\theta(t)$ is the step function. The first term of $a(t,T)$ can drive the phase transition thermally and the second through the effect of the optical pump. For simplicity, we assumed that the laser modifies the free energy instantaneously upon incidence and that the latter relaxes back to equilibrium in a time $\tau_0$. The order parameter does not follow the abrupt change of $L$, but instead is reduced on a characteristic time $\tau_s$. This reflects the fact that just after the pump absorption the number density of broken Cooper pairs is of the order of the photon density of the laser ($\sim 10^{-7}$ formula unit$^{-1}$, for $\Omega = 1$ mJ/cm$^3$), which is negligible compared to the density of broken Cooper pairs at the end of the suppression process ($\sim 10^{-5}$ formula unit$^{-1}$, for $\Omega = 1$ mJ/cm$^3$ [‡]). Additional details of the simulations are provided in Supplemental Note 3 [30]. We note that the only adjustable parameter in the modeling was the conversion factor $k$.

Representative numerical solutions of the time-dependent Ginzburg-Landau equation for various laser energies are displayed in Fig. 4(a). The suppression time of such transients is plotted in Fig. 4(b) together with the quantity $\Delta\psi/\psi_{\text{eq}} = (\psi_{\text{eq}} - \min[\psi(t)])/\psi_{\text{eq}}$, where $\psi_{\text{eq}}$ is the order parameter at equilibrium. In close resemblance to the experiments, $\tau_s$ was highest when $\Delta\psi/\psi_{\text{eq}}$ virtually saturated, namely when the order parameter—and the Cooper pair condensate—was nearly destroyed. Further, the simulated $\tau_s(\Omega)$ curve also exhibited a lower tail for $\Omega > \Omega_{\text{sat}}$ than in the $\Omega < \Omega_{\text{sat}}$ region due to the pronounced steepness of the Landau potential attainable at high fluences. In the phenomenon of critical slowing-down, the temperature quenches the order parameter and, consequently, substantially reduces its rate of change [18].

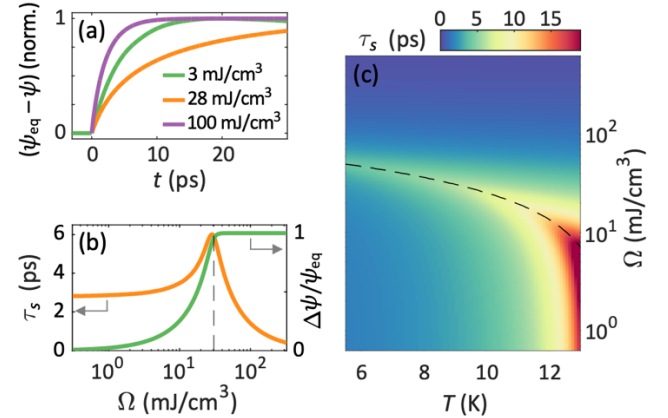

FIG. 4. Ginzburg-Landau modeling. (a) Solutions of the time-dependent Ginzburg-Landau equation for different fluences. (b) Superconductivity suppression time juxtaposed with the maximum changes of the order parameter ($\Delta\psi/\psi_{\text{eq}}$) simulated for $T = 9$ K. The dashed line marks the fluence for which $\psi$ was reduced by 95% relative to its initial value ($\Delta\psi/\psi_{\text{eq}}(9\,K) = 0.95$). (c) Suppression time as a function of temperature and fluence. The dashed curve indicates the fluences for which $\Delta\psi/\psi_{\text{eq}}(T) = 0.95$.

Our results reveal that a similar effect can be induced in a highly non-equilibrium scenario by fully depleting the superconducting condensate via an ultrashort optical excitation on a timescale comparable to that of dynamical fluctuations [50].

Finally, we repeated the calculation at various ambient temperatures and obtained the $T$, $\Omega$ phase map shown in Fig. 4(c), where the suppression time was maximized at the temperature-dependent threshold fluence required to eradicate $\psi$ (dashed line). Despite the qualitative agreement with Fig. 3(c), it is worth noting that the simulations yielded greater $\tau_s$ values close to $\Omega_{\text{sat}}(T)$ than the experiments. This partly stemmed from the sparse selection of experimental fluences which were not sufficiently close to $\Omega_{\text{sat}}(T)$. Moreover, the discrepancies may also have been influenced by spatial inhomogeneities across the NbN film given that such variations weaken the acuity of the phenomenon of critical slowing-down [18]. The inhomogeneities may be connected with the spatial dependence of the equilibrium properties of the sample or with the coexistence of normal and superconducting regions during the superconductivity quenching process [51].

In summary, we have shown that the ultrafast optical suppression of superconductivity in NbN displays a marked protraction when the system is dynamically pushed to the phase transition boundary. Notably, this far-from-equilibrium counterpart to critical slowing-down cannot be accounted for by thermodynamic criticality, since the effect occurs on a timescale comparable to–or, sufficiently close to $T_c$, even faster than [50]–that of superconducting fluctuations. Nevertheless, our results can be adequately described with a phenomenological extension of the time-dependent Ginzburg-Landau theory for non-equilibrium systems. This work establishes OPTP spectroscopy as a powerful method to investigate superconducting dynamics driven by extreme perturbations. More broadly, it highlights the unique suitability of pump-probe techniques to directly detect critical behavior in real time and motivates the search of further signatures of universality away from equilibrium, for instance in pseudogap [26], metal-insulator [52], multiferroic [53] or nematic [54] phase transitions, where conventional critical slowing-down has been observed.

Acknowledgements— We thank Karl Bergson Hallberg and Jacob Linder for fruitful discussions. C.C. acknowledges the Royal Society Research Fellowship, the Leverhulme Trust Research Project grant (Grant No. RPG-2023-271), and the UKRI Frontier Research Guarantee grant (Grant No. EP/Z000637/1). This project has received funding from the European Union's Horizon 2020 Research and Innovation Programme under the Marie Skłodowska-Curie Grant Agreement No. 861300 (COMRAD) and from the Winton Programme for the Physics of Sustainability. Additionally, this work was partly supported by the France 2030 Programme "Lorraine Initiative of Excellence" (reference ANR-15-IDEX-04-LUE), by the France 2030 government investment plan managed by the French National Research Agency PEPR SPIN – SUPERSPIN (grant reference ANR-22-EXSP-0012), by the Norwegian Research Council funding scheme Centers of Excellence, "QuSpin", and by the COST Action CA21144 "Superconducting Nanodevices and Quantum Materials for Coherent Manipulation".

**End Matter**

*Appendix A: Sample fabrication*—The NbN thin film was deposited by magnetron sputtering at room temperature and with a DC power of 165 W, using a Nb target in a mixed atmosphere of Ar and N2 with flow rates of 30 sccm and 3.25 sccm, respectively. The full structure of the sample is MgO(100) (substrate)/MgO(5 nm)/NbN(15 nm)/MgO(5 nm)/Pt(2 nm), where the MgO/Pt capping was included to protect the superconducting film. The thickness of the layers was confirmed through X-ray reflectivity measurements.

*Appendix B: Experimental setup*—The THz TDS and OPTP experiments were implemented using a titanium-sapphire laser amplifier which produced 35-fs-long pulses with a central wavelength of 800 nm at a repetition rate of 5 kHz. The laser output was employed to generate THz pulses by optical rectification in a 1 mm thick, (110)-oriented ZnTe crystal. The THz radiation was transmitted through the sample and detected via electro-optic sampling in a second, identical ZnTe crystal with the aid of gate pulses emitted by the titanium-sapphire laser and delayed with respect to the THz pulses by a variable time $t_{\mathrm{THz}}$. In the OPTP measurements, the pump also originated from the same laser system, and we employed the experimental geometry represented in Fig. 7 of reference [33].

*Appendix C: Fitting of the superconductivity quenching dynamics*—We extracted the timescale $(\tau_s)$ of the superconductivity suppression transients by fitting the OPTP signal to the function

$$\frac{\Delta S(t)}{\Delta S_s} = \frac{1}{\xi - 2}\left[\xi + 2 + -\frac{4}{1 - K\exp(-t/\tau_s)}\right] \quad (2)$$

which has the form of the solution of the RT model for the photo-excited QP population $(n)$ in the pre-phonon bottleneck regime [15]. In that context, $\xi$ and $K$ are dimensionless parameters defined in [15] in terms of the Cooper pair breaking $(\beta)$ and recombination $(R)$ rates and the populations of QPs and HFPs just after the pump arrival. In the weak perturbation limit, $n$ is proportional to $\Delta S$ [12,17], giving a physical justification to the use of Eq. (2). For higher fluences, we simply regarded Eq. (2) as a fit function that enabled a systematic determination of the superconductivity suppression time.

At the lowest measured temperature of 5.5 K and in the $\Omega \ll \Omega_{\mathrm{sat}}$ regime, we fitted the $\tau_s$ data in Fig. 3(a) to the relation $\tau_s = [\beta^2/4 + 2R\beta(\Omega/\Delta)]^{-1/2}$ [5] from the RT theory, taking $\beta$ and $R$ as adjustable parameters. The fitted values of $\beta = 0.44 \pm 0.04$ ps$^{-1}$ and $R = 117 \pm 9$ (ps × conventional unit cell)$^{-1}$ are of the same order of magnitude as those reported in the literature [9], which adds further credence to our method of analysis based on the OPTP signal.